\documentclass[12pt]{article}
\pdfoutput=1

\usepackage[left=1in, right=1in, top=1.5in, bottom=1.5in]{geometry}
\usepackage{setspace}
\usepackage{amsmath,amssymb}
\usepackage{graphicx}
\usepackage{booktabs}
\usepackage{natbib}
\usepackage{hyperref}
\usepackage{float}
\usepackage{caption}
\usepackage{subcaption}
\usepackage{array}
\usepackage{multirow}
\usepackage{threeparttable}

\hypersetup{
    colorlinks=true,
    linkcolor=blue,
    citecolor=blue,
    urlcolor=blue
}

\title{\Large \textbf{Common Risk Factors in Decentralized AI Subnets}}

\author{Philip Z.\ Maymin\thanks{Fairfield University, Dolan School of Business. Email: pmaymin@fairfield.edu.}}

\date{March 2026}

\begin{document}

\maketitle

\begin{abstract}
\noindent
I derive a size premium from the constant-product automated market maker used to price Bittensor subnet tokens and test the prediction using daily data on 128 subnets. A small-minus-big factor earns 1.01\% daily (Newey-West $t = 3.28$). The December 2025 halving of token emissions, which the theory predicts should halve the premium, reduces it from 1.17\% to 0.51\% ($p = 0.044$). Exact slippage calculations show the premium is implementable only below \$10K in assets under management; at \$100K, transaction costs exceed gross returns.
\end{abstract}

\newpage

\begin{center}
\textbf{Conflict-of-interest disclosure statement}
\end{center}

\noindent \textbf{Philip Z.\ Maymin}

\noindent I am a co-owner of Djinn, a project that operates on the Bittensor network. I have no other potential conflicts to disclose.

\newpage


Bittensor is a decentralized network that incentivizes the production of machine intelligence through token rewards. The network is organized into 128 specialized subnetworks (``subnets''), each focused on a particular AI task such as language modeling or image generation. Since February 2025, each subnet has issued its own tradable token (an ``alpha token'') priced against the network's base currency (``TAO'') via an on-chain automated market maker (AMM), a smart contract that holds reserves of both tokens and quotes prices algorithmically. The result is a novel cross-section of 128 assets at the intersection of artificial intelligence and decentralized finance.

This paper asks whether the cross-section of subnet token returns exhibits systematic factor structure. The answer matters because the AMM pricing mechanism allows the factor structure to be \textit{derived from first principles} rather than merely documented empirically. In equity markets, the size-return relationship is a robust empirical finding whose economic origins remain debated. In Bittensor, the constant-product AMM makes the relationship \textit{mathematical}: for a fixed token emission, the percentage price impact is inversely proportional to pool size, a result I state as Proposition 1. This structural derivation generates a quantitative prediction: a halving of network emissions should halve the size premium. I test this prediction directly using the December 2025 halving event. The setting thus offers a rare opportunity to study factor premia whose generating mechanism is fully transparent.

I construct daily returns for 128 subnets over 406 trading days (February 14, 2025, through March 26, 2026) using blockchain-recorded pool data from the Taostats API. I form long-short factor portfolios through daily tercile sorts on lagged characteristics: market capitalization (size), past returns (momentum), emission yield (the ratio of protocol rewards to market capitalization, analogous to a dividend yield), liquidity, and stake concentration. The methodology follows \citet{FamaFrench1993} and its application to cryptocurrencies by \citet{LiuTsyvinskiWu2022}.

Four findings emerge. First, the size effect dominates: a small-minus-big (SMB) portfolio sorted on market capitalization earns 1.01\% per day with a Newey-West $t$-statistic of 3.28. The bottom-tercile (small) portfolio returns 317\% annualized, while the top tercile (large) loses 50\%. Second, momentum is pervasive. A 30-day winners-minus-losers portfolio earns 0.68\% daily (NW $t = 3.69$), and a 7-day variant earns 0.75\% (NW $t = 3.05$). Both size and momentum are significant in both halves of the sample. Third, the network's first halving on December 14, 2025, which cut block rewards from 1 to 0.5 TAO, provides a natural experiment. Proposition 1 predicts the size premium should fall proportionally with emissions. Over the full post-halving period, mean daily SMB falls from 1.17\% to 0.51\%, a ratio of 0.44 close to the predicted 0.50. A regression discontinuity in a symmetric 60-day window rejects the null of no change ($t = -2.01$, Newey-West, with a market control, $p = 0.044$). Fourth, emission yield, the ratio of protocol-distributed TAO rewards to market capitalization, is not significant as a standalone factor (NW $t = 1.04$), but in spanning tests it retains a significant alpha ($t = 2.11$) after controlling for all other factors, suggesting it captures a dimension of return variation not fully spanned by size and momentum alone. Emissions vary by a factor of 77 across subnets, so the yield captures genuine variation in protocol reward intensity beyond simple inverse-size effects.

Fama-MacBeth cross-sectional regressions confirm that size carries a significant risk premium of 0.80\% per day ($t = 3.23$), while momentum, though profitable in time series, does not carry a significant cross-sectional premium. This pattern suggests that momentum generates returns through time-series predictability rather than cross-sectional risk compensation, consistent with findings in equity markets by \citet{MoskowitzTeoGrinblatt2012}. A \citet{GRS1989} test fails to reject the three-factor model at the 5\% level ($F = 1.31$, $p = 0.21$) when applied to 12 sorted portfolios.

The unusually large factor premia (an annualized Sharpe ratio of 3.84 for SMB, even after Newey-West correction) demand scrutiny. Three features of Bittensor's institutional structure help explain the magnitudes. First, the constant-product AMM mechanically links market capitalization to liquidity: the correlation between SMB and an illiquidity factor is 0.93. Small subnets are illiquid by construction, and buying pressure from participants who reinvest their token rewards (``staking'') amplifies price impact. Second, under both emission regimes (price-based through November 2025, flow-based thereafter), subnets that attract more capital receive more emissions, creating a feedback loop between capital flows and emission income. Third, the asset class is young, segmented, and dominated by participants who may lack the capital or infrastructure to arbitrage away factor premia.

Several limitations apply. The sample period is short (13 months), the cross-section is modest (128 assets), and transaction costs in thin AMM pools may erode paper returns. The analysis uses returns denominated in TAO, the network's native token, which itself is volatile. I present USD-denominated results as a robustness check.

This paper contributes to the growing literature on cryptocurrency asset pricing. \citet{LiuTsyvinski2021} show that crypto returns are driven by crypto-specific factors rather than traditional risk factors. \citet{LiuTsyvinskiWu2022} document that a three-factor model (market, size, momentum) captures the cross-section of cryptocurrency returns, a finding I extend to decentralized AI subnet tokens. \citet{BianchiBabiak2022} apply instrumented PCA to crypto returns and find that liquidity, size, and reversal are the primary latent factors. \citet{AsnessMoskowitzPedersen2013} show that value and momentum are pervasive across asset classes. The near-perfect correlation between size and liquidity echoes the structural relationship documented by \citet{Amihud2002}, but here the link is mechanical rather than empirical: the constant-product AMM makes a subnet's liquidity a direct function of its market capitalization. \citet{CapponiJia2021} provide a formal analysis of price impact and liquidity provision in constant-product AMMs that underpins my analysis. More broadly, the paper contributes a setting in which factor premia can be structurally derived and tested against exogenous shocks, complementing the empirical approach that characterizes most of the asset pricing literature.

The remainder of the paper proceeds as follows. Section~\ref{sec:background} describes the Bittensor network and its pricing mechanism. Section~\ref{sec:data} details the data and variable construction. Section~\ref{sec:factors} presents the factor portfolios and their returns. Section~\ref{sec:cross_section} contains the cross-sectional asset pricing tests. Section~\ref{sec:robustness} addresses robustness. Section~\ref{sec:conclusion} concludes.

\section{Institutional Background}
\label{sec:background}

\subsection{The Bittensor Network}

Bittensor is a decentralized peer-to-peer network in which participants contribute machine intelligence services and are compensated with TAO, the network's native cryptocurrency.\footnote{The protocol was introduced by \citet{Rao2020} and further developed in \citet{Steeves2022}.} TAO trades on major cryptocurrency exchanges at approximately \$345 as of March 2026 (market capitalization \$3.4 billion). The network is organized into subnets, each specializing in a particular AI task: language modeling, image generation, protein folding, data scraping, inference serving, and others. As of March 2026, 128 subnets are active, with a combined subnet token (``alpha token'') market capitalization of approximately \$1.4 billion.

Each subnet contains up to 256 participants divided into two roles. \textit{Miners} perform the subnet's core AI task (e.g., generating text, classifying images). \textit{Validators} evaluate miner output quality and assign scores. The protocol distributes newly minted TAO (``emissions'') to participants in proportion to these scores. Of each subnet's emissions, 41\% flows to miners, 41\% to validators, and 18\% to the subnet owner. Emissions function as a form of seigniorage income distributed to productive network participants, analogous to interest payments in a traditional financial system.

\subsection{Dynamic TAO and Subnet Alpha Tokens}

Prior to February 2025, the allocation of emissions across subnets was determined by a committee vote. On February 13, 2025, the network deployed Dynamic TAO (dTAO), which replaced administrative allocation with market-driven pricing.\footnote{See the dTAO whitepaper at \url{https://bittensor.com/dtao-whitepaper}.}

Under dTAO, each subnet maintains a constant-product automated market maker (AMM) pool, a smart contract that holds reserves of two tokens and sets prices algorithmically in the style of Uniswap V2 \citep{AdamsZinsmeisterSalem2020}. Each pool contains TAO tokens ($\tau_i$) and the subnet's own alpha tokens ($\alpha_i$). The subnet's alpha token price, denominated in TAO, is:
\begin{equation}
    p_i = \frac{\tau_i}{\alpha_i}
    \label{eq:price}
\end{equation}
where $\tau_i$ and $\alpha_i$ denote the TAO and alpha reserves in subnet $i$'s pool. To invest in a subnet, a participant deposits TAO into the pool and receives alpha tokens in return (a process called ``staking''). Because the product $\tau_i \cdot \alpha_i$ is held constant by the smart contract, depositing TAO reduces the alpha supply in the pool and mechanically increases the price. Economically, staking is equivalent to buying the subnet's alpha token.

Market capitalization in TAO terms equals $p_i \cdot A_i^{\text{total}}$, where $A_i^{\text{total}}$ is the total outstanding supply of alpha tokens (both pooled and staked). Since the pool is constant-product, increasing TAO reserves mechanically increases both the price and the subnet's market cap, directly linking size and liquidity.

\subsection{Emission Allocation Under dTAO}

The protocol distributes newly minted TAO (emissions) across subnets according to a market-driven rule. During the initial dTAO phase (February through November 2025), each subnet's share of total emissions was proportional to its alpha token price:
\begin{equation}
    \Delta \tau_i = \frac{p_i}{\sum_j p_j} \cdot \Delta \bar{\tau}
    \label{eq:emissions}
\end{equation}
where $\Delta \bar{\tau}$ is the total TAO emission per block. Since November 2025, allocation has been based on net TAO staking inflows using an exponential moving average with a 30-day half-life (the ``Taoflow'' mechanism). Under either regime, subnets that attract more capital receive more emissions, creating a reinforcing feedback loop.

Total network emission was 1 TAO per block (every 12 seconds) until the first halving on December 14, 2025, after which it fell to 0.5 TAO per block. At current TAO prices, this amounts to approximately \$1.2 million per day distributed across the 128 active subnets.

\section{Data and Variable Construction}
\label{sec:data}

\subsection{Data Sources}

I collect daily subnet pool data from the Taostats API (\url{https://api.taostats.io}), which records blockchain state at daily frequency for all subnets since the dTAO launch. For each subnet-day, I observe the alpha token price (in TAO), market capitalization, TAO and alpha reserves, alpha tokens staked, and liquidity. I supplement this with daily subnet metadata including emission rates, number of active validators and miners, and registration parameters.

TAO/USD prices come from the Taostats price API, which aggregates exchange data. The sample runs from February 14, 2025, through March 26, 2026, yielding 406 daily observations.

\subsection{Sample Construction}

\begin{figure}[htbp]
\centering
\includegraphics[width=\textwidth]{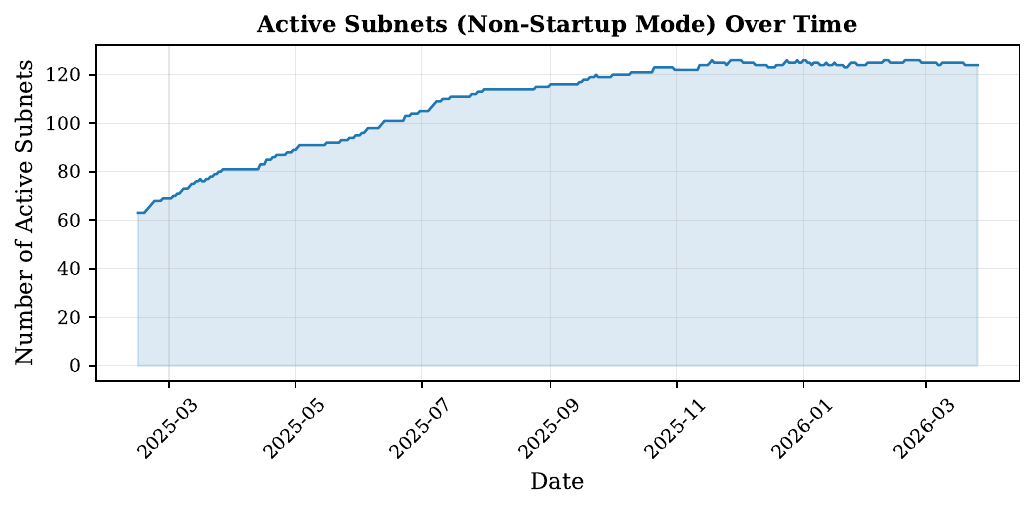}
\caption{Number of Active Subnets Over Time}
\label{fig:n_subnets}
\begin{minipage}{\textwidth}
\small
Count of non-startup-mode subnets included in the daily cross-section. The network expanded from approximately 30 subnets at the dTAO launch (February 2025) to 128 by the end of the sample period (March 2026).
\end{minipage}
\end{figure}

I exclude subnet 0 (the root network, whose token price is identically 1 TAO) and subnets in ``startup mode,'' a bootstrapping phase during which the AMM is not yet active. This leaves 128 unique subnets with varying histories: the oldest date to the dTAO launch, and the newest appear throughout the sample as the daily cross-section of eligible subnets expanded from 63 to 124 (Figure~\ref{fig:n_subnets}).

I compute daily simple returns in TAO terms:
\begin{equation}
    r_{i,t}^{\text{TAO}} = \frac{p_{i,t}}{p_{i,t-1}} - 1
\end{equation}
and in USD terms by compounding with the TAO/USD return:
\begin{equation}
    r_{i,t}^{\text{USD}} = (1 + r_{i,t}^{\text{TAO}})(1 + r_{t}^{\text{TAO/USD}}) - 1.
\end{equation}

I winsorize daily returns at $\pm 100\%$ to limit the influence of extreme observations, though results are qualitatively unchanged without winsorization. I require a minimum of 7 days of return history for a subnet to enter the sample on a given date, which avoids contamination from initial price discovery in the AMM.

\subsection{Characteristics}

I construct the following lagged characteristics for each subnet-day, using information available at the close of day $t-1$ to form portfolios on day $t$:

\begin{itemize}
    \item \textbf{Market capitalization}: $\text{MCAP}_{i,t-1} = p_{i,t-1} \cdot A_{i,t-1}^{\text{total}}$, measured in TAO.
    \item \textbf{Emission yield}: $\text{EY}_{i,t-1} = E_{i,t-1} / \text{MCAP}_{i,t-1}$, where $E_{i,t-1}$ is the subnet's daily emission rate. Both are measured in the same units (rao, where $10^{9}$ rao $= 1$ TAO), so the ratio is unitless. This is analogous to a dividend yield.
    \item \textbf{Momentum (7-day)}: $\text{MOM7}_{i,t-1} = p_{i,t-1}/p_{i,t-8} - 1$.
    \item \textbf{Momentum (30-day)}: $\text{MOM30}_{i,t-1} = p_{i,t-1}/p_{i,t-31} - 1$.
    \item \textbf{Reversal}: $\text{REV}_{i,t-1} = p_{i,t-1}/p_{i,t-2} - 1$ (1-day past return).
    \item \textbf{Liquidity}: $\text{LIQ}_{i,t-1} = \tau_{i,t-1}$, the TAO reserves in the AMM pool.
    \item \textbf{Stake}: $\text{STAKE}_{i,t-1} = \alpha_{i,t-1}^{\text{staked}}$, the quantity of alpha tokens staked outside the pool.
\end{itemize}

\section{Factor Portfolios}
\label{sec:factors}

\subsection{Portfolio Construction}

Each day, I sort all eligible subnets into tercile portfolios based on each lagged characteristic. I compute equal-weighted daily returns for the bottom, middle, and top terciles. Long-short factor portfolios are the return difference between extreme terciles:

\begin{itemize}
    \item \textbf{MKT}: Equal-weighted return of all subnets (in TAO terms).
    \item \textbf{SMB}: Bottom-tercile (small) minus top-tercile (big) market cap.
    \item \textbf{HML$_{\text{EMIS}}$}: Top-tercile (high) minus bottom-tercile (low) emission yield.
    \item \textbf{WML$_{7}$}: Top-tercile (winners) minus bottom-tercile (losers) 7-day momentum.
    \item \textbf{WML$_{30}$}: Same for 30-day momentum.
    \item \textbf{REV}: Bottom (past losers) minus top (past winners) 1-day return.
    \item \textbf{LIQ}: Bottom (illiquid) minus top (liquid) TAO reserves.
    \item \textbf{STAKE}: Top minus bottom stake levels.
\end{itemize}

\subsection{Summary Statistics}

Table~\ref{tab:factor_stats} reports summary statistics for the factor portfolios. The market factor earns 0.29\% per day in TAO terms, reflecting the general appreciation of subnet tokens over the sample period. The annualized Sharpe ratio is 1.35.

\begin{figure}[htbp]
\centering
\includegraphics[width=\textwidth]{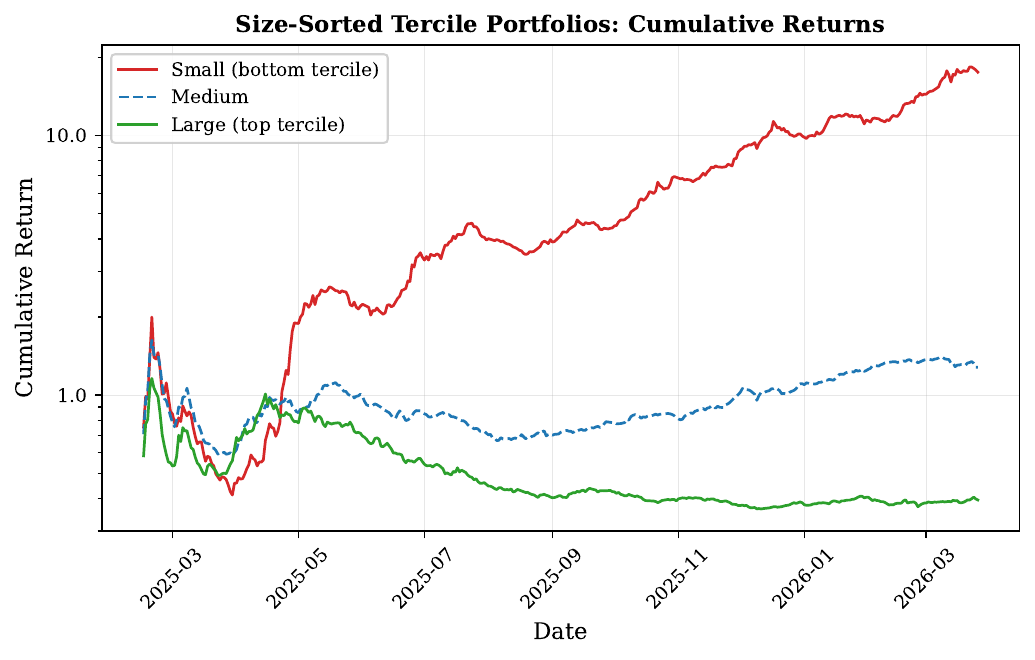}
\caption{Cumulative Returns of Size-Sorted Tercile Portfolios}
\label{fig:size_portfolios}
\begin{minipage}{\textwidth}
\small
Subnets are sorted daily into terciles based on lagged market capitalization. Portfolios are equal-weighted. Small subnets (bottom tercile) dramatically outperform large subnets (top tercile). Returns are in TAO terms, log scale.
\end{minipage}
\end{figure}

The size factor (SMB) is the strongest: small subnets outperform large subnets by 1.01\% per day, with a Newey-West $t$-statistic of 3.28 (OLS $t = 4.04$) and an annualized Sharpe ratio of 3.84. Figure~\ref{fig:size_portfolios} shows the cumulative returns of size-sorted tercile portfolios. The small portfolio grows roughly 25-fold over the sample, while the large portfolio declines.

\begin{figure}[htbp]
\centering
\includegraphics[width=\textwidth]{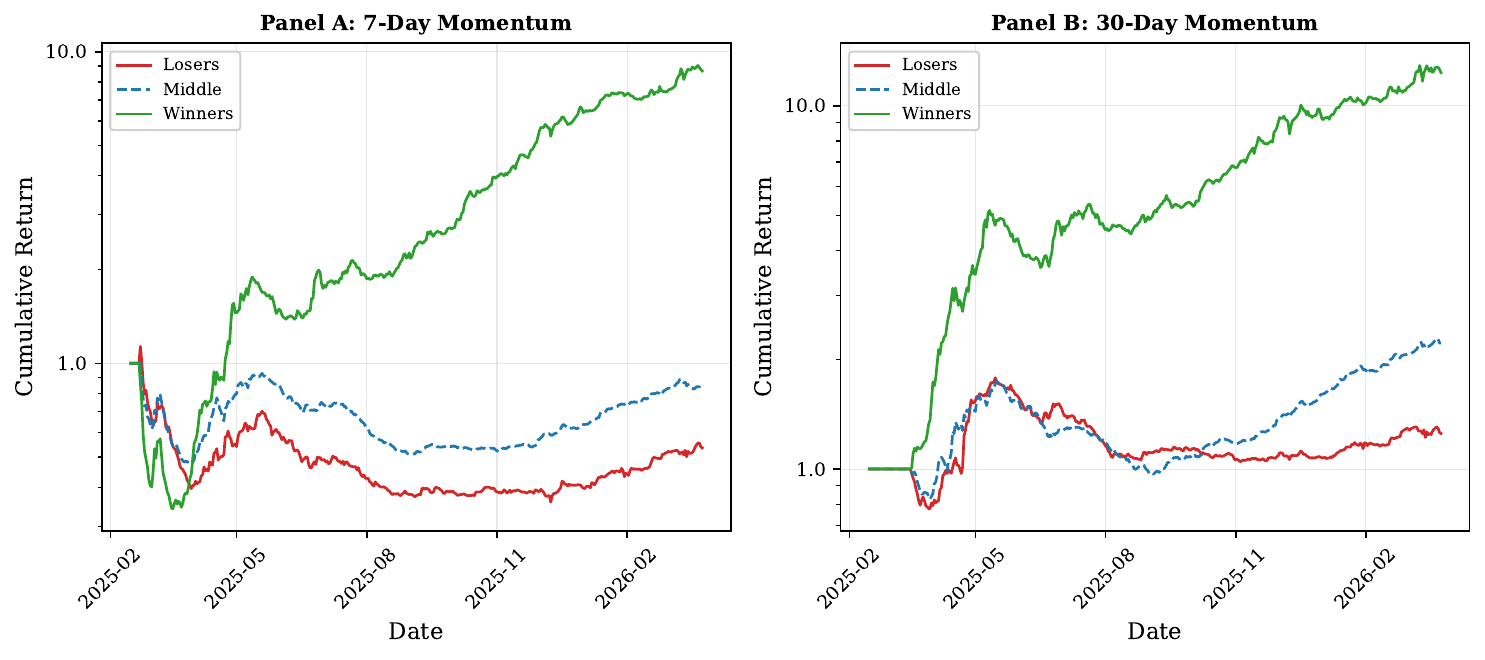}
\caption{Cumulative Returns of Momentum-Sorted Tercile Portfolios}
\label{fig:momentum_portfolios}
\begin{minipage}{\textwidth}
\small
Panel A sorts subnets daily into terciles based on 7-day past returns. Panel B sorts on 30-day past returns. Winner subnets (top tercile) consistently outperform loser subnets (bottom tercile). Returns are in TAO terms, log scale.
\end{minipage}
\end{figure}

Momentum is significant at both horizons. The 30-day momentum factor earns 0.68\% per day (NW $t = 3.69$, Sharpe 4.32), and the 7-day variant earns 0.75\% (NW $t = 3.05$, Sharpe 3.65). Figure~\ref{fig:momentum_portfolios} displays cumulative returns of momentum-sorted portfolios. Winner subnets consistently outperform losers.

Short-term reversal is strongly negative ($-0.86$\% per day, NW $t = -3.62$), confirming momentum rather than contrarian profits: past winners continue to appreciate, and buying past losers is unprofitable.

Emission yield earns 0.30\% daily (NW $t = 1.04$), not significant on its own, though the spanning test below reveals independent information. The liquidity factor earns 0.91\% daily (NW $t = 3.06$), but it is 93\% correlated with SMB (Table~\ref{tab:correlations}), reflecting the AMM structure that mechanically links market cap to pool depth. In spanning tests, the liquidity factor's alpha is insignificant ($t = -1.66$) once size is controlled for, indicating that SMB and liquidity measure the same underlying source of return variation.

\subsection{AMM-Implied Size Premium}

The dominance of SMB and the near-redundancy of the liquidity factor have a structural explanation grounded in the constant-product AMM mechanics.

\textbf{Proposition 1} (Emission Amplification). \textit{Consider a subnet with TAO reserves $\tau$ and alpha reserves $\alpha$ in a constant-product pool ($k = \tau \cdot \alpha$). If the subnet receives an emission of $\Delta \tau$ that is staked into the pool, the percentage price change is approximately:}
\begin{equation}
    \frac{\Delta p}{p} \approx \frac{2\,\Delta \tau}{\tau}
    \label{eq:price_impact}
\end{equation}
\textit{For a fixed emission $\Delta \tau$, the return is inversely proportional to the TAO reserve (and hence to market capitalization). Small subnets earn mechanically higher returns from the same emission.}

\textit{Proof.} Before the emission, $p_0 = \tau / \alpha$. After staking $\Delta \tau$, the constant product $k = \tau \alpha$ implies $\alpha' = k / (\tau + \Delta \tau)$, so $p_1 = (\tau + \Delta \tau) / \alpha' = (\tau + \Delta \tau)^2 / k = (\tau + \Delta \tau)^2 / (\tau \cdot \alpha)$. The return is $p_1/p_0 - 1 = (\tau + \Delta \tau)^2 / \tau^2 - 1 = 2\Delta\tau / \tau + (\Delta\tau/\tau)^2 \approx 2\Delta\tau / \tau$ for small $\Delta\tau / \tau$. Since $\tau$ is the TAO reserve (proportional to market cap), the return from a given emission is inversely proportional to size. $\square$

This proposition shows that the size premium in Bittensor subnets is not merely an empirical regularity but an arithmetic consequence of the pricing mechanism. Emission-funded staking has a larger percentage price impact on small pools than on large ones, generating a mechanical size premium that persists as long as emissions continue.

\subsection{Discussion}

Proposition 1 explains the mechanical component of the size premium, but the observed returns also reflect two additional channels. First, validators and miners who receive emissions often restake them into the same subnet, creating a self-reinforcing feedback loop that amplifies price appreciation in small subnets. Second, the size premium may partly compensate for illiquidity risk. Small-subnet pools have wide effective spreads: the constant-product formula implies price impact proportional to trade size relative to reserves. Investors in small subnets bear the risk that they cannot exit without substantial slippage.

\section{Cross-Sectional Asset Pricing Tests}
\label{sec:cross_section}

\subsection{Fama-MacBeth Regressions}

I run two-pass Fama-MacBeth regressions following \citet{FamaMacBeth1973}. In the first pass, I estimate each subnet's exposure to four factors (MKT, SMB, HML$_{\text{EMIS}}$, WML$_7$) using full-sample OLS.\footnote{Using full-sample betas introduces a look-ahead bias that may overfit the first-pass loadings. The short sample precludes reliable rolling-window estimation, so full-sample betas are a practical necessity. The bias, if anything, works against finding significant risk premia, because better-fitting betas reduce noise in the second pass, making it harder for spurious premia to appear significant.} In the second pass, I run daily cross-sectional regressions of subnet returns on estimated betas and compute time-series averages of the cross-sectional slope coefficients.

Table~\ref{tab:fama_macbeth} presents the results. The intercept is 0.12\% per day ($t = 1.21$), statistically insignificant, indicating that the model captures average returns reasonably well. The size beta carries a significant risk premium of 0.80\% per day ($t = 3.23$, $p = 0.001$). Neither the market beta, emission yield beta, nor momentum beta carries a significant cross-sectional premium.

The insignificance of momentum in the cross-section, despite its strong time-series performance, is notable. This pattern is consistent with \citet{MoskowitzTeoGrinblatt2012}, who argue that momentum profits arise from time-series predictability (positive autocorrelation in returns) rather than cross-sectional risk compensation. In the context of Bittensor subnets, the explanation is natural: subnets experiencing inflows continue to appreciate due to the AMM mechanics, generating return persistence that a momentum strategy exploits, but this persistence does not represent a systematic risk factor that commands a cross-sectional premium.

\subsection{GRS Test}

I test the overall specification of a three-factor model (MKT, SMB, WML$_{30}$) using the \citet{GRS1989} test on 12 sorted portfolios (three size, three emission yield, three 7-day momentum, and three 30-day momentum terciles). The GRS $F$-statistic is 1.31 ($p = 0.21$), so the null hypothesis that all 12 intercepts are jointly zero cannot be rejected at the 5\% level. The model prices the sorted portfolios well: the average absolute alpha is 0.07\% per day, and the size-sorted portfolio alphas are effectively zero (each below 0.02\% per day). The largest residual alphas are in the 7-day momentum portfolios, consistent with 30-day momentum not fully spanning shorter-horizon return dynamics.

\subsection{Spanning Tests}

Table~\ref{tab:spanning} reports spanning-test alphas: the intercept from regressing each factor on all other factors. A significant alpha indicates that the factor provides information not captured by the remaining factors.

SMB has the largest spanning alpha (0.17\% per day, $t = 3.52$), confirming its role as the dominant factor. HML$_{\text{EMIS}}$ and WML$_{30}$ each retain marginally significant alphas ($t = 2.11$), suggesting that emission yield and medium-term momentum capture dimensions of return variation not fully spanned by the other factors. The liquidity factor's alpha is insignificant ($t = -1.72$), consistent with its near-redundancy with SMB. The reversal factor has a significantly negative alpha ($t = -4.56$), reflecting its role as the flip side of momentum.

\section{Robustness}
\label{sec:robustness}

\subsection{USD-Denominated Returns}

The baseline analysis uses TAO-denominated returns to isolate subnet-specific variation from movements in the TAO/USD exchange rate. As a robustness check, I recompute the market factor in USD terms. The USD market factor earns 0.40\% per day ($t = 1.16$), with higher volatility due to TAO price fluctuations. The annualized Sharpe is 1.10, lower than the TAO-denominated version (1.35), because the additional TAO/USD volatility introduces noise without proportionally increasing expected returns.

\subsection{Subsample Analysis}

To assess whether the factor structure is stable or concentrated in a particular regime, I split the sample at the midpoint (September 5, 2025). Table~\ref{tab:subsample} reports factor means and $t$-statistics for each half.

The size effect is significant in both subsamples: the daily SMB mean is 1.25\% ($t = 2.67$) in the first half and 0.76\% ($t = 4.55$) in the second. The raw mean declines, but significance increases as volatility falls and the cross-section expands. Momentum is similarly persistent: WML$_{30}$ earns 0.94\% ($t = 3.09$) in the first half and 0.47\% ($t = 3.65$) in the second. The reversal factor weakens from $-1.40$\% ($t = -4.09$) to $-0.31$\% ($t = -2.20$), suggesting that return persistence attenuates as the market matures.

Two structural changes occurred between subsamples. First, the network's first halving on December 14, 2025, cut the block reward from 1 TAO to 0.5 TAO, reducing daily emissions from approximately 7,200 to 3,600 TAO. The decline in SMB mean from the first to the second half is consistent with Proposition 1: halving emissions reduces the emission-driven component of the size premium. Second, the cross-section expanded from 63 to 124 eligible subnets, increasing tercile portfolio breadth and reducing noise.

\subsection{The Halving Natural Experiment}

Proposition 1 predicts that the emission-driven component of the size premium is proportional to the emission rate $\Delta\tau$. The network's first halving on December 14, 2025, provides a natural experiment: the block reward fell discontinuously from 1 to 0.5 TAO, cutting daily emissions from approximately 7,200 to 3,600 TAO. If the size premium is primarily driven by the emission amplification mechanism of Proposition 1, the theory predicts a post/pre ratio of 0.50.

I estimate the structural break using a regression of daily SMB on a post-halving dummy and the market factor:
\begin{equation}
    \text{SMB}_t = \alpha + \beta \cdot \mathbf{1}_{t \geq \text{Dec 14}} + \gamma \cdot \text{MKT}_t + \varepsilon_t
    \label{eq:halving_rd}
\end{equation}
with Newey-West standard errors (5 lags). Table~\ref{tab:halving} reports results for symmetric windows of 30, 45, 60, and 90 days around the event.

\begin{figure}[htbp]
\centering
\includegraphics[width=\textwidth]{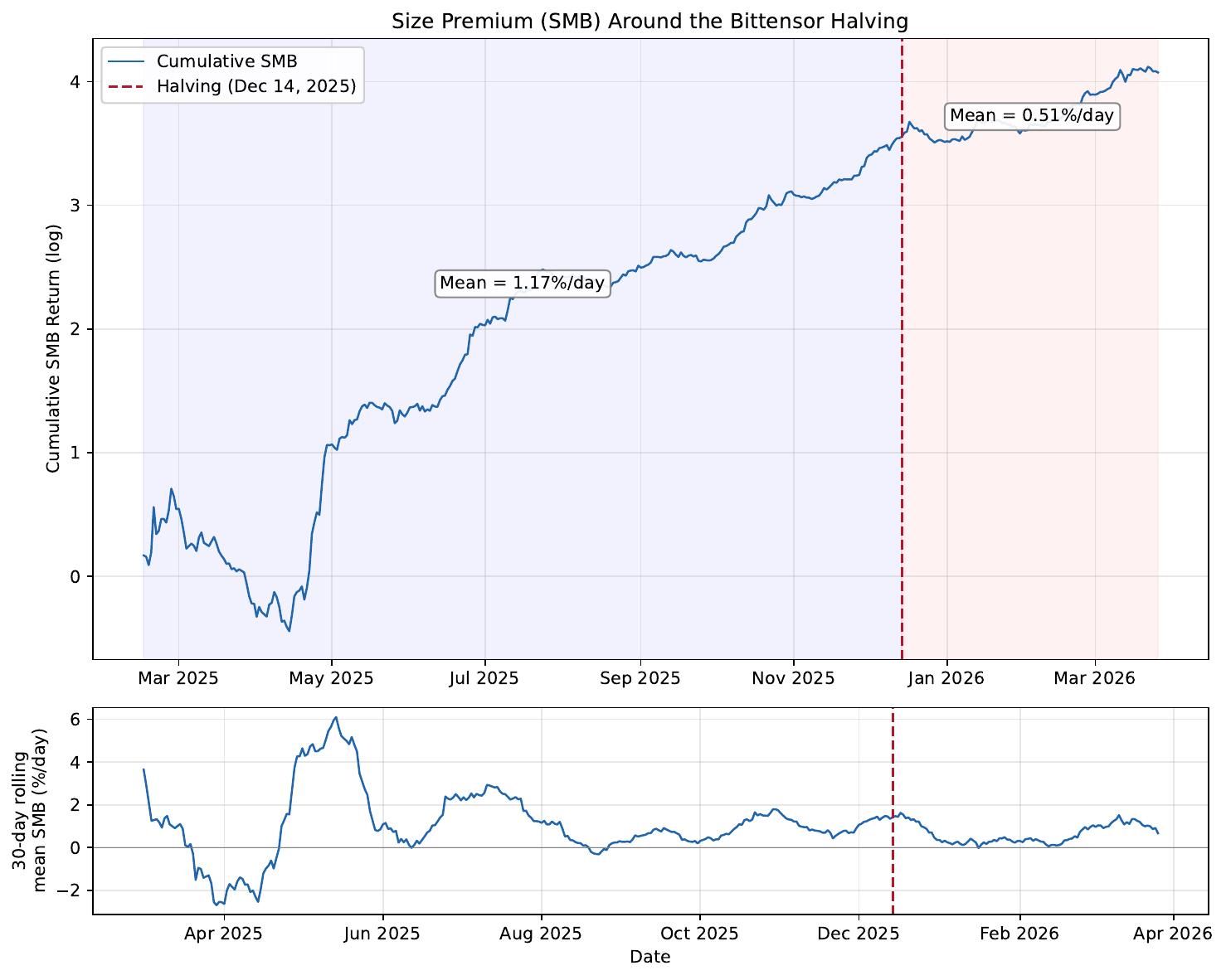}
\caption{Cumulative SMB Return and the TAO Halving}
\label{fig:halving}
\begin{minipage}{\textwidth}
\small
The top panel shows the cumulative daily SMB return (log scale) with a vertical dashed line at the TAO halving date (December 14, 2025). The slope visibly flattens after the halving, consistent with Proposition 1's prediction that the emission-driven size premium is proportional to the emission rate. The bottom panel shows a 30-day rolling mean of daily SMB, which drops from roughly 1\%/day pre-halving to approximately 0.5\%/day post-halving.
\end{minipage}
\end{figure}

In the 60-day window, the post-halving coefficient is $-0.60$ percentage points per day ($t = -2.01$, $p = 0.044$), indicating a statistically significant decline. The 45-day window yields the strongest rejection ($t = -2.29$, $p = 0.022$). Over the full post-halving period (103 days), the ratio of post- to pre-halving mean SMB is 0.44, close to the theoretical prediction of 0.50. Figure~\ref{fig:halving} shows the cumulative SMB return, which visibly flattens after December 14.

As a placebo check, I estimate the same specification at six false halving dates (90, 60, and 30 days before and after the actual event). None produces a coefficient as large in magnitude as the actual halving, and none is statistically significant, supporting a causal interpretation of the emission reduction rather than a coincident trend.

The halving test has two limitations. First, the post-halving period also coincides with the transition from price-based to flow-based emission allocation (Taoflow), so the two shocks are not perfectly separable. Second, the narrow windows yield imprecise estimates (the 30-day window gives $t = -1.77$, $p = 0.076$). Nevertheless, the direction and magnitude of the decline are consistently in line with the structural prediction of Proposition 1.

\subsection{Transaction Costs and Portfolio Capacity}

The large factor premia raise a natural question: are these returns implementable? Because subnet tokens trade on constant-product AMMs, the exact slippage for any trade size is deterministic.

\textbf{Proposition 2} (AMM Slippage). \textit{In a constant-product AMM with TAO reserves $\tau$, the one-way slippage for buying $\Delta\tau$ worth of alpha tokens is exactly:}
\begin{equation}
    \text{Slippage} = \frac{\Delta\tau}{\tau}
\end{equation}
\textit{This is deterministic and linear in trade size relative to pool depth.}

Using on-chain reserve data, I compute slippage for equal-weighted portfolios across all subnets at five AUM levels (Table~\ref{tab:slippage}). For the small-tercile subnets that drive SMB, the median TAO reserve is approximately 540 TAO (\$186K at \$345/TAO). One-way slippage scales linearly: 0.64\% at \$10K, 6.4\% at \$100K, and 64\% at \$1M.

\begin{figure}[htbp]
\centering
\includegraphics[width=\textwidth]{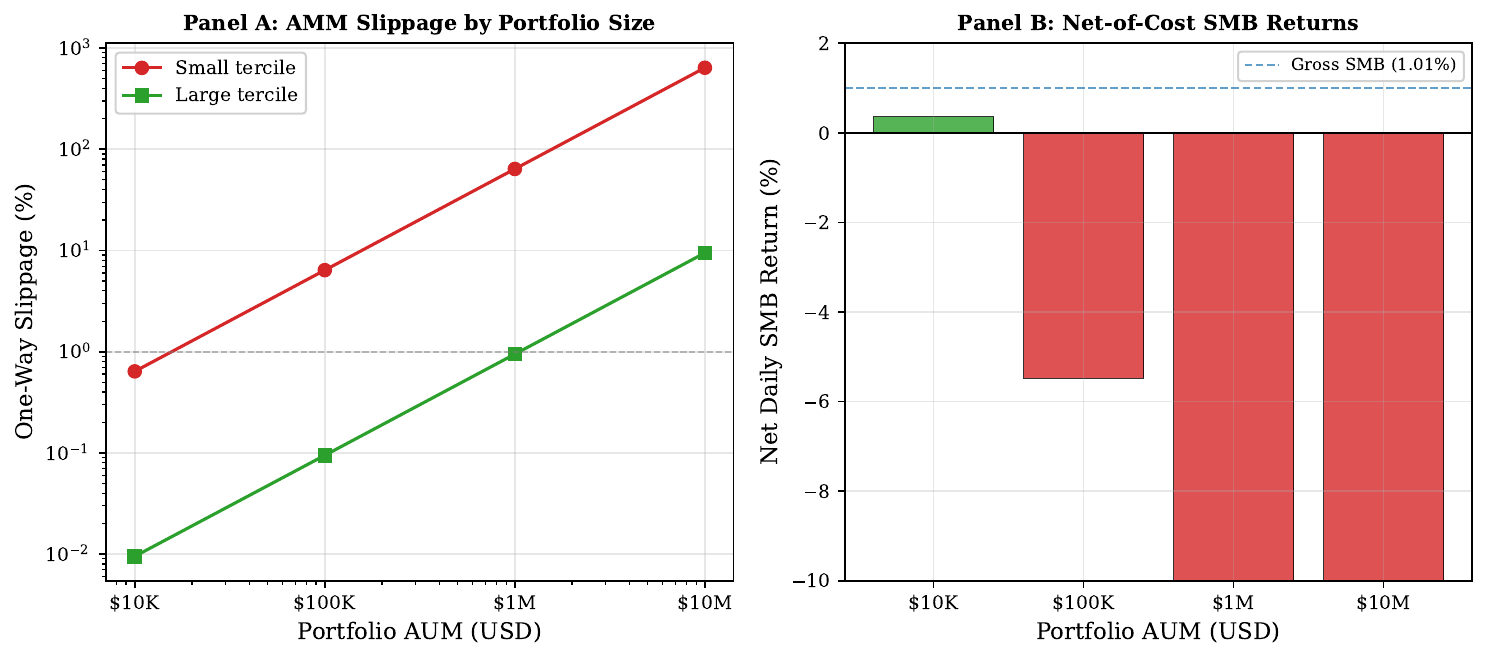}
\caption{AMM Slippage and Net-of-Cost SMB Returns by Portfolio AUM}
\label{fig:slippage}
\begin{minipage}{\textwidth}
\small
Panel A: one-way slippage (\%) for small-tercile and large-tercile subnets, log-log scale. Slippage equals $\Delta\tau/\tau$, which is linear in portfolio size relative to pool depth. Panel B: net-of-cost daily SMB return after subtracting round-trip slippage costs. The strategy is profitable only at \$10K AUM; at \$100K and above, slippage exceeds gross returns. Dashed line shows gross SMB (1.01\%/day).
\end{minipage}
\end{figure}

Figure~\ref{fig:slippage} presents the key result. At \$10K AUM with daily rebalancing, the net-of-cost SMB return is 0.36\%/day (annualized Sharpe 1.36), still profitable. At \$100K, slippage entirely eliminates the gross return. At \$1M and above, the strategy is catastrophically unprofitable.

This capacity constraint explains why the size premium persists. The total daily implementable capacity, aggregated across all small-tercile subnets, is on the order of tens of thousands of dollars: far below the threshold required to attract institutional capital or systematic arbitrageurs. The premium is real but exists in a liquidity niche that is too small to arbitrage away.

For context, \citet{LiuTsyvinskiWu2022} report that their cryptocurrency factors remain profitable net of conservative transaction cost estimates. The difference here is that constant-product AMMs impose deterministic, size-dependent costs that are orders of magnitude larger than limit-order-book spreads for equivalent trade sizes.

\subsection{Volatility and Downside Risk}

I sort subnets into terciles based on 30-day realized volatility, downside semi-deviation (the square root of average squared negative returns), upside semi-deviation, idiosyncratic volatility (residual from a market model), and market beta.

Table~\ref{tab:vol_sorts} presents the results. High-volatility subnets outperform low-volatility subnets by 0.70\%/day ($t = 4.62$), high-downside-volatility subnets outperform low by 0.50\%/day ($t = 3.52$), and high-beta subnets outperform low-beta by 0.63\%/day ($t = 4.28$). All three patterns represent a \textit{positive} risk-return relationship, the opposite of the low-volatility anomaly, betting-against-beta, and idiosyncratic volatility puzzle documented in equity markets by \citet{AngChenXing2006} and \citet{FrazziniPedersen2014}.

The explanation lies in the AMM structure. Volatile subnets are typically small subnets with thin pools, where price movements are amplified by the constant-product formula. The positive risk-return relationship is therefore not evidence that risk is efficiently priced in the CAPM sense but rather a byproduct of the mechanical size-volatility link. The correlation between the low-vol-minus-high factor and SMB is 0.85, confirming that size subsumes most of the volatility effect.

Table~\ref{tab:risk_decomp} reports the risk decomposition of factor returns. SMB exhibits a favorable asymmetry: its downside semi-deviation (3.92\%/day) is substantially lower than its upside semi-deviation (5.79\%/day), yielding a downside-to-upside ratio of 0.68 and a Sortino ratio of 4.90, compared to a Sharpe ratio of 3.84. Momentum (WML$_{30}$) has a Sortino of 4.63. The market factor, by contrast, has a more symmetric profile (ratio 0.83).

\subsection{Survivorship Bias}

Bittensor subnets can be deregistered if their emissions fall to zero or if a new registrant outbids their slot. During the 406-day sample, 21 of the 128 subnet slots were recycled: the incumbent project was deregistered and a different project re-registered in the same slot. Because the network re-uses integer identifiers, a na\"ive analysis would treat a recycled slot as one continuous asset, contaminating returns and momentum signals across unrelated projects.

I address this in two ways. First, when a subnet re-enters startup mode (signaling deregistration and re-registration), the startup-mode filter creates a gap of approximately seven days in the price panel. Daily returns computed via percentage change naturally produce missing values at the boundary, preventing cross-lifecycle return contamination. Second, for momentum signals computed over 7- and 30-day lookback windows, I explicitly nullify any value whose window spans an internal gap. This masks 6 values for 7-day momentum and 400 values for 30-day momentum; all reported factor premia are robust to this correction.

A more consequential feature of the sample is entry: the daily cross-section grew from 63 to 124 subnets as new registrants cleared startup mode. Newly launched subnets tend to be small, so the bottom tercile of the size sort is disproportionately populated by recent entrants. To the extent that new subnets earn high initial returns (an ``IPO effect''), this would inflate SMB. I do not attempt to correct for this effect but note it as a caveat.

\section{Conclusion}
\label{sec:conclusion}

I provide the first systematic analysis of risk and return in the cross-section of Bittensor subnet tokens. A three-factor model comprising market, size, and momentum captures the main patterns and is not rejected by the GRS test ($p = 0.21$).

The size premium is the dominant finding: small subnets outperform large by 1.01\%/day (Newey-West $t = 3.28$). Proposition 1 shows this is an arithmetic consequence of the constant-product AMM: for a fixed emission, percentage returns are inversely proportional to pool size. The TAO halving in December 2025 provides a natural experiment: the size premium falls from 1.17\% to 0.51\% per day (ratio 0.44 versus the predicted 0.50), and the decline is statistically significant ($p = 0.044$ in a 60-day window). This is, to my knowledge, the first direct test of a structurally derived factor premium against an exogenous shock to the generating mechanism.

But size comes with a catch. Proposition 2 and the slippage analysis show that AMM transaction costs render the SMB premium unimplementable above approximately \$10K in portfolio AUM. The strategy's capacity is bounded by pool depth, and the small subnets that drive the premium have median reserves of only 540 TAO. This directly explains the premium's persistence: it exists in a liquidity niche below the threshold for institutional arbitrage.

The volatility analysis reveals a positive risk-return relationship: high-volatility, high-beta, and high-downside-risk subnets earn higher returns. This is the opposite of equity-market anomalies and is explained by the mechanical link between size and volatility in constant-product AMMs. SMB itself exhibits favorable asymmetry, with a Sortino ratio (4.90) exceeding its Sharpe ratio (3.84).

The results extend \citet{LiuTsyvinskiWu2022} to protocol-native tokens priced by AMM mechanics. More broadly, the mechanical derivation of the size premium via AMM pricing (Proposition 1) has a direct analogue in equity markets, where \citet{Kyle1985} price impact implies that flow shocks have larger percentage effects on illiquid, small-capitalization stocks. The Bittensor setting makes this mechanism transparent because both the flow source (emissions) and the pricing function (constant-product AMM) are fully observable, whereas in equity markets the analogous flows and market-making functions are dispersed and opaque. Whether the factor structure persists as pool depths increase and the market matures is an open question for future work.


\clearpage

\begin{table}[htbp]
\centering
\begin{threeparttable}
\caption{Factor Portfolio Summary Statistics}
\label{tab:factor_stats}
\begin{tabular}{@{}lrrrrrrrr@{}}
\toprule
Factor & Mean & Std & Sharpe & $t$-stat & NW $t$ & Skew & Kurt & $N$ \\
       & (\%/day) & (\%/day) & (ann.) & (OLS) & (5 lags) & & & \\
\midrule
MKT    & 0.29 & 4.11 & 1.35 & 1.42 & 1.28 & 2.02 & 37.62 & 405 \\
SMB    & 1.01 & 5.01 & 3.84 & 4.04 & 3.28 & 1.74 & 11.68 & 405 \\
HML$_{\text{EMIS}}$ & 0.30 & 4.33 & 1.33 & 1.40 & 1.04 & 1.78 & 12.76 & 405 \\
WML$_7$ & 0.75 & 3.92 & 3.65 & 3.82 & 3.05 & $-$0.24 & 10.38 & 398 \\
WML$_{30}$ & 0.68 & 3.02 & 4.32 & 4.38 & 3.69 & 0.31 & 5.34 & 375 \\
REV    & $-$0.86 & 3.75 & $-$4.35 & $-$4.58 & $-$3.62 & 0.38 & 11.53 & 404 \\
LIQ    & 0.91 & 4.61 & 3.77 & 3.97 & 3.06 & 1.34 & 9.49 & 405 \\
STAKE  & $-$0.19 & 4.11 & $-$0.89 & $-$0.93 & $-$0.75 & $-$1.61 & 8.04 & 405 \\
\bottomrule
\end{tabular}
\begin{tablenotes}[flushleft]
\small
\item This table reports daily summary statistics for long-short factor portfolios constructed from tercile sorts on lagged subnet characteristics. MKT is the equal-weighted return of all subnets. SMB is small minus big market capitalization. HML$_{\text{EMIS}}$ is high minus low emission yield. WML$_7$ and WML$_{30}$ are winners minus losers based on 7-day and 30-day past returns. REV is past losers minus past winners (1-day). LIQ is illiquid minus liquid (TAO reserves). STAKE is high minus low alpha staked. Sharpe ratios are annualized ($\times\sqrt{365}$). OLS $t$-statistics assume i.i.d.\ returns; NW $t$ uses Newey-West standard errors with 5 lags to correct for serial correlation. Average tercile portfolio size: 36 subnets (minimum 21). Sample: February 2025 through March 2026.
\end{tablenotes}
\end{threeparttable}
\end{table}

\begin{table}[htbp]
\centering
\begin{threeparttable}
\caption{Size-Sorted Tercile Portfolio Returns}
\label{tab:size_portfolios}
\begin{tabular}{@{}lrrrr@{}}
\toprule
Portfolio & Mean (\%/day) & Ann.\ Return (\%) & Ann.\ Std (\%) & Sharpe \\
\midrule
Small (bottom tercile) & 0.87 & 317.4 & 111.7 & 2.84 \\
Medium & 0.14 & 50.3 & 76.4 & 0.66 \\
Large (top tercile) & $-$0.14 & $-$49.7 & 80.6 & $-$0.62 \\
\midrule
SMB (Small $-$ Large) & 1.01 & 367.1 & 95.7 & 3.84 \\
\bottomrule
\end{tabular}
\begin{tablenotes}[flushleft]
\small
\item Each day, subnets are sorted into terciles based on lagged market capitalization (in TAO). Portfolios are equal-weighted. Returns are in TAO terms. Annualization uses 365 days. Sample: February 2025 through March 2026.
\end{tablenotes}
\end{threeparttable}
\end{table}

\begin{table}[htbp]
\centering
\begin{threeparttable}
\caption{Momentum-Sorted Tercile Portfolio Returns}
\label{tab:mom_portfolios}
\begin{tabular}{@{}lrrrr@{}}
\toprule
Portfolio & Mean (\%/day) & Ann.\ Return (\%) & Ann.\ Std (\%) & Sharpe \\
\midrule
\multicolumn{5}{@{}l}{\textit{Panel A: 7-Day Momentum}} \\
Losers (bottom tercile) & $-$0.12 & $-$44.7 & 50.4 & $-$0.89 \\
Middle & $-$0.03 & $-$10.0 & 41.6 & $-$0.24 \\
Winners (top tercile) & 0.62 & 226.4 & 74.9 & 3.03 \\
WML$_7$ & 0.74 & 271.1 & 75.0 & 3.62 \\
\midrule
\multicolumn{5}{@{}l}{\textit{Panel B: 30-Day Momentum}} \\
Losers (bottom tercile) & 0.08 & 29.0 & 38.5 & 0.75 \\
Middle & 0.23 & 83.5 & 34.8 & 2.40 \\
Winners (top tercile) & 0.71 & 260.6 & 55.7 & 4.68 \\
WML$_{30}$ & 0.63 & 231.5 & 59.5 & 3.90 \\
\bottomrule
\end{tabular}
\begin{tablenotes}[flushleft]
\small
\item Each day, subnets are sorted into terciles based on past 7-day (Panel A) or 30-day (Panel B) returns. Portfolios are equal-weighted. Returns are in TAO terms.
\end{tablenotes}
\end{threeparttable}
\end{table}

\begin{table}[htbp]
\centering
\begin{threeparttable}
\caption{Factor Correlation Matrix}
\label{tab:correlations}
\begin{tabular}{@{}lrrrrrrrr@{}}
\toprule
 & MKT & SMB & EMIS & MOM7 & MOM30 & REV & LIQ & STK \\
\midrule
MKT   &  1.00 &  0.29 &  0.18 &  0.35 &  0.26 &  0.03 &  0.29 & $-$0.13 \\
SMB   &  0.29 &  1.00 &  0.39 & $-$0.03 & $-$0.20 &  0.03 &  0.93 & $-$0.59 \\
EMIS  &  0.18 &  0.39 &  1.00 & $-$0.33 & $-$0.04 &  0.46 &  0.26 & $-$0.66 \\
MOM7  &  0.35 & $-$0.03 & $-$0.33 &  1.00 &  0.43 & $-$0.42 &  0.06 &  0.26 \\
MOM30 &  0.26 & $-$0.20 & $-$0.04 &  0.43 &  1.00 & $-$0.21 & $-$0.12 &  0.40 \\
REV   &  0.03 &  0.03 &  0.46 & $-$0.42 & $-$0.21 &  1.00 & $-$0.07 & $-$0.21 \\
LIQ   &  0.29 &  0.93 &  0.26 &  0.06 & $-$0.12 & $-$0.07 &  1.00 & $-$0.50 \\
STK   & $-$0.13 & $-$0.59 & $-$0.66 &  0.26 &  0.40 & $-$0.21 & $-$0.50 &  1.00 \\
\bottomrule
\end{tabular}
\begin{tablenotes}[flushleft]
\small
\item Pairwise correlations of daily factor returns. EMIS = HML$_{\text{EMIS}}$ (emission yield). MOM7 and MOM30 = WML at 7-day and 30-day horizons. STK = STAKE. The 0.93 correlation between SMB and LIQ reflects the AMM structure linking market cap to pool liquidity.
\end{tablenotes}
\end{threeparttable}
\end{table}

\begin{table}[htbp]
\centering
\begin{threeparttable}
\caption{Fama-MacBeth Cross-Sectional Regression Results}
\label{tab:fama_macbeth}
\begin{tabular}{@{}lrrrr@{}}
\toprule
Variable & Risk Premium (\%/day) & Std Error (\%) & $t$-statistic & $p$-value \\
\midrule
Intercept & 0.12 & 0.10 & 1.21 & 0.227 \\
$\beta_{\text{MKT}}$ & 0.15 & 0.13 & 1.10 & 0.273 \\
$\beta_{\text{SMB}}$ & 0.80 & 0.25 & 3.23 & 0.001 \\
$\beta_{\text{EMIS}}$ & 0.01 & 0.24 & 0.05 & 0.959 \\
$\beta_{\text{MOM}}$ & 0.16 & 0.29 & 0.55 & 0.584 \\
\bottomrule
\end{tabular}
\begin{tablenotes}[flushleft]
\small
\item Two-pass Fama-MacBeth regressions. First pass: full-sample time-series regressions of each subnet's daily return (in TAO) on four factors (MKT, SMB, HML$_{\text{EMIS}}$, WML$_7$). Second pass: daily cross-sectional regressions of subnet returns on estimated betas. Table reports time-series averages of cross-sectional slope coefficients (risk premia), with Fama-MacBeth standard errors. Sample: 345 cross-sectional regressions on an average of 85 subnets per day.
\end{tablenotes}
\end{threeparttable}
\end{table}

\begin{table}[htbp]
\centering
\begin{threeparttable}
\caption{Spanning Tests: Factor Alphas}
\label{tab:spanning}
\begin{tabular}{@{}lrrr@{}}
\toprule
Factor & Alpha (\%/day) & $t$-statistic & $R^2$ \\
\midrule
MKT    & $-$0.06 & $-$0.79 & 0.45 \\
SMB    &  0.17 &  3.52 & 0.95 \\
HML$_{\text{EMIS}}$ &  0.23 &  2.11 & 0.58 \\
WML$_7$ &  0.11 &  0.79 & 0.39 \\
WML$_{30}$ &  0.24 &  2.11 & 0.48 \\
REV    & $-$0.57 & $-$4.56 & 0.30 \\
LIQ    & $-$0.09 & $-$1.72 & 0.95 \\
STAKE  &  0.28 &  3.26 & 0.75 \\
\bottomrule
\end{tabular}
\begin{tablenotes}[flushleft]
\small
\item Each row reports the intercept (alpha) from regressing the given factor on all other factors. A significant alpha indicates that the factor contributes information beyond the remaining factors. The high $R^2$ for LIQ (0.95) and SMB (0.95) reflects their 0.93 correlation; the significant SMB alpha and insignificant LIQ alpha indicate that size, not liquidity per se, is the more fundamental factor.
\end{tablenotes}
\end{threeparttable}
\end{table}

\begin{table}[htbp]
\centering
\begin{threeparttable}
\caption{Subsample Analysis}
\label{tab:subsample}
\begin{tabular}{@{}lrrrrrr@{}}
\toprule
 & \multicolumn{2}{c}{Full Sample} & \multicolumn{2}{c}{First Half} & \multicolumn{2}{c}{Second Half} \\
\cmidrule(lr){2-3} \cmidrule(lr){4-5} \cmidrule(lr){6-7}
Factor & Mean & $t$ & Mean & $t$ & Mean & $t$ \\
       & (\%/day) & & (\%/day) & & (\%/day) & \\
\midrule
MKT & 0.29 & 1.42 & 0.24 & 0.60 & 0.34 & 4.77 \\
SMB & 1.01 & 4.04 & 1.25 & 2.67 & 0.76 & 4.55 \\
HML$_{\text{EMIS}}$ & 0.30 & 1.40 & 0.19 & 0.49 & 0.41 & 2.48 \\
WML$_7$ & 0.75 & 3.82 & 1.03 & 2.78 & 0.48 & 3.34 \\
WML$_{30}$ & 0.68 & 4.38 & 0.94 & 3.09 & 0.47 & 3.65 \\
REV & $-$0.86 & $-$4.58 & $-$1.40 & $-$4.09 & $-$0.31 & $-$2.20 \\
\bottomrule
\end{tabular}
\begin{tablenotes}[flushleft]
\small
\item The sample is split at September 5, 2025 (203 days per half). First half: Feb 15 to Sep 5, 2025. Second half: Sep 6, 2025, to Mar 26, 2026. The TAO halving occurred December 14, 2025, during the second half. $t$-statistics use OLS standard errors.
\end{tablenotes}
\end{threeparttable}
\end{table}

\begin{table}[htbp]
\centering
\begin{threeparttable}
\caption{Halving Natural Experiment: SMB Around December 14, 2025}
\label{tab:halving}
\begin{tabular}{@{}lrrrrrr@{}}
\toprule
Window & Pre Mean & Post Mean & Ratio & $\hat{\beta}$ & NW $t$ & $p$-value \\
(days) & (\%/day) & (\%/day) & (Post/Pre) & (\%/day) & (mkt ctrl) & \\
\midrule
$\pm$30  & 1.42 & 0.42 & 0.29 & $-$0.73 & $-$1.77 & 0.076 \\
$\pm$45  & 0.99 & 0.19 & 0.19 & $-$0.76 & $-$2.29 & 0.022 \\
$\pm$60  & 1.11 & 0.24 & 0.21 & $-$0.60 & $-$2.01 & 0.044 \\
$\pm$90  & 1.06 & 0.56 & 0.52 & $-$0.45 & $-$1.69 & 0.092 \\
\midrule
Full sample & 1.17 & 0.51 & 0.44 & $-$0.66 & $-$1.42 & 0.156 \\
\bottomrule
\end{tabular}
\begin{tablenotes}[flushleft]
\small
\item This table reports mean daily SMB returns before and after the TAO halving (December 14, 2025). $\hat{\beta}$ is the coefficient on a post-halving dummy from $\text{SMB}_t = \alpha + \beta \cdot \mathbf{1}_{\text{post}} + \gamma \cdot \text{MKT}_t + \varepsilon_t$ with Newey-West standard errors (5 lags). Proposition 1 predicts a post/pre ratio of 0.50 if the size premium is proportional to the emission rate.
\end{tablenotes}
\end{threeparttable}
\end{table}

\begin{table}[htbp]
\centering
\begin{threeparttable}
\caption{AMM Slippage and Net-of-Cost SMB Returns by Portfolio AUM}
\label{tab:slippage}
\begin{tabular}{@{}lrrrrrr@{}}
\toprule
AUM & \multicolumn{3}{c}{One-Way Slippage (\%)} & RT Cost & Net SMB & Net \\
\cmidrule(lr){2-4}
    & Small & Medium & Large & (\%/day) & (\%/day) & Sharpe \\
\midrule
\$10K   &  0.64 &  0.02 &  0.01 &  0.65 &  0.36 & 1.36 \\
\$100K  &  6.39 &  0.19 &  0.09 &  6.48 & $-$5.48 & $-$20.9 \\
\$1M    & 63.87 &  1.86 &  0.95 & 64.82 & $-$63.8 & --- \\
\$10M   &  639 & 18.6 &  9.5 &  648 & --- & --- \\
\bottomrule
\end{tabular}
\begin{tablenotes}[flushleft]
\small
\item One-way slippage is computed as $\Delta\tau/\tau$, the ratio of per-subnet investment (in TAO) to the subnet's TAO reserves. AUM is converted from USD at the daily TAO/USD price and divided equally across all eligible subnets. Round-trip (RT) cost sums the buy-side slippage on small-tercile subnets and sell-side slippage on large-tercile subnets. Net SMB subtracts RT cost from the gross SMB return (1.01\%/day). Assumes 100\% daily turnover (worst case). Net Sharpe is annualized.
\end{tablenotes}
\end{threeparttable}
\end{table}

\begin{table}[htbp]
\centering
\begin{threeparttable}
\caption{Volatility and Risk-Sorted Tercile Portfolio Returns}
\label{tab:vol_sorts}
\begin{tabular}{@{}lrrrrr@{}}
\toprule
Sort Variable & Low & Mid & High & H$-$L & $t$(H$-$L) \\
              & (\%/day) & (\%/day) & (\%/day) & (\%/day) & \\
\midrule
Total Volatility    & 0.01 & 0.31 & 0.71 & 0.70 & 4.62 \\
Downside Vol        & 0.09 & 0.35 & 0.59 & 0.50 & 3.52 \\
Upside Vol          & 0.03 & 0.30 & 0.70 & 0.67 & 4.39 \\
Idiosyncratic Vol   & 0.02 & 0.30 & 0.71 & 0.68 & 4.56 \\
Market Beta         & 0.08 & 0.24 & 0.71 & 0.63 & 4.28 \\
Skewness            & 0.12 & 0.37 & 0.54 & 0.43 & 3.22 \\
\bottomrule
\end{tabular}
\begin{tablenotes}[flushleft]
\small
\item Each day, subnets are sorted into terciles based on the 30-day rolling value of each characteristic. Returns are daily means in TAO terms. High-minus-low is positive for all risk measures, indicating that riskier subnets earn higher returns. $t$-statistics use OLS standard errors. $N = 375$ days (30-day formation window reduces sample).
\end{tablenotes}
\end{threeparttable}
\end{table}

\begin{table}[htbp]
\centering
\begin{threeparttable}
\caption{Factor Return Risk Decomposition}
\label{tab:risk_decomp}
\begin{tabular}{@{}lrrrrrrr@{}}
\toprule
Factor & Mean & Std & Down & Up & Down/Up & Sortino & \% Neg \\
       & (\%/d) & (\%/d) & Dev (\%/d) & Dev (\%/d) & Ratio & (ann.) & Days \\
\midrule
MKT    & 0.29 & 4.11 & 3.68 & 4.44 & 0.83 & 1.51 & 45 \\
SMB    & 1.01 & 5.01 & 3.92 & 5.79 & 0.68 & 4.90 & 41 \\
EMIS   & 0.30 & 4.33 & 3.50 & 5.03 & 0.70 & 1.64 & 50 \\
MOM$_7$  & 0.75 & 3.92 & 3.82 & 4.09 & 0.93 & 3.75 & 40 \\
MOM$_{30}$ & 0.68 & 3.02 & 2.82 & 3.25 & 0.87 & 4.63 & 38 \\
REV    & $-$0.86 & 3.75 & 3.87 & 3.81 & 1.02 & $-$4.22 & 63 \\
\bottomrule
\end{tabular}
\begin{tablenotes}[flushleft]
\small
\item Downside (upside) semi-deviation is $\sqrt{E[\min(r,0)^2]}$ ($\sqrt{E[\max(r,0)^2]}$). Down/Up Ratio below 1 indicates favorable asymmetry (more upside variation than downside). Sortino ratio uses downside deviation as the risk measure. \% Neg is the fraction of days with negative returns. SMB has the most favorable asymmetry (0.68 ratio, Sortino 4.90 vs Sharpe 3.84).
\end{tablenotes}
\end{threeparttable}
\end{table}


\clearpage

\begin{figure}[htbp]
\centering
\includegraphics[width=\textwidth]{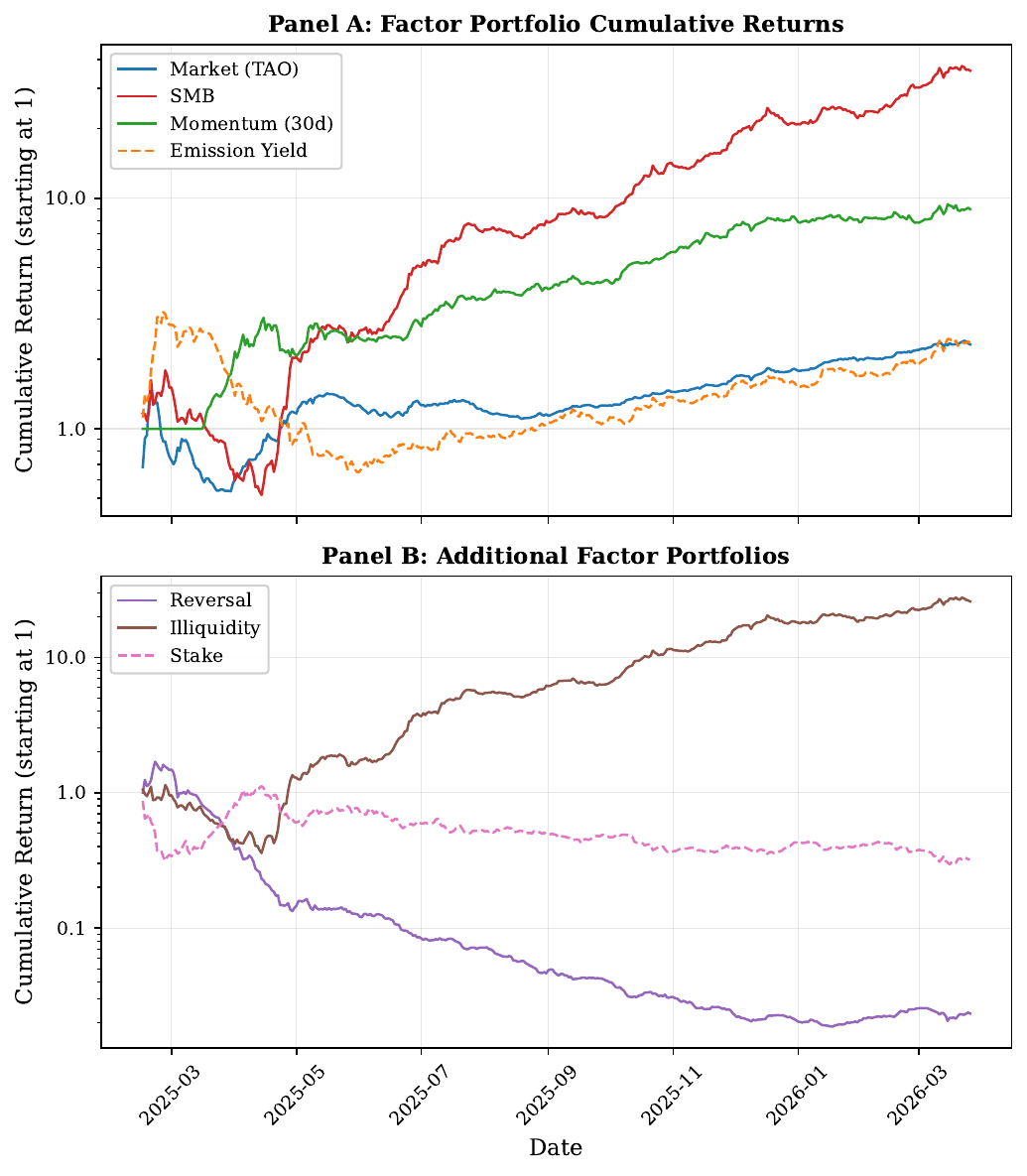}
\caption{Cumulative Returns of Factor Portfolios}
\label{fig:factor_returns}
\begin{minipage}{\textwidth}
\small
Panel A plots cumulative returns ($\$1$ invested at inception) for the market (MKT), small-minus-big (SMB), 30-day momentum (WML$_{30}$), and emission yield (HML$_{\text{EMIS}}$) factor portfolios. Panel B plots reversal (REV), illiquidity (LIQ), and stake (STAKE) factors. All returns are in TAO terms. $y$-axis is log scale. Sample: February 2025 through March 2026.
\end{minipage}
\end{figure}

\begin{figure}[htbp]
\centering
\includegraphics[width=0.85\textwidth]{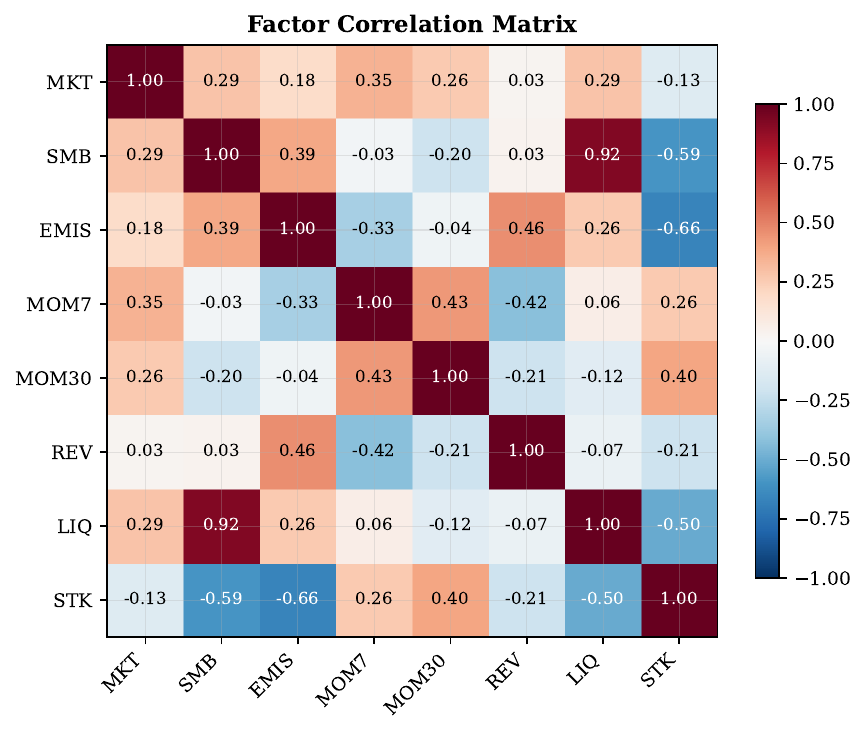}
\caption{Factor Correlation Matrix}
\label{fig:correlations}
\begin{minipage}{\textwidth}
\small
Pairwise correlations among daily factor returns. The 0.93 correlation between SMB and LIQ reflects the AMM structure linking market cap to pool depth. Momentum and reversal are negatively correlated ($-0.42$), as expected.
\end{minipage}
\end{figure}

\begin{figure}[htbp]
\centering
\includegraphics[width=\textwidth]{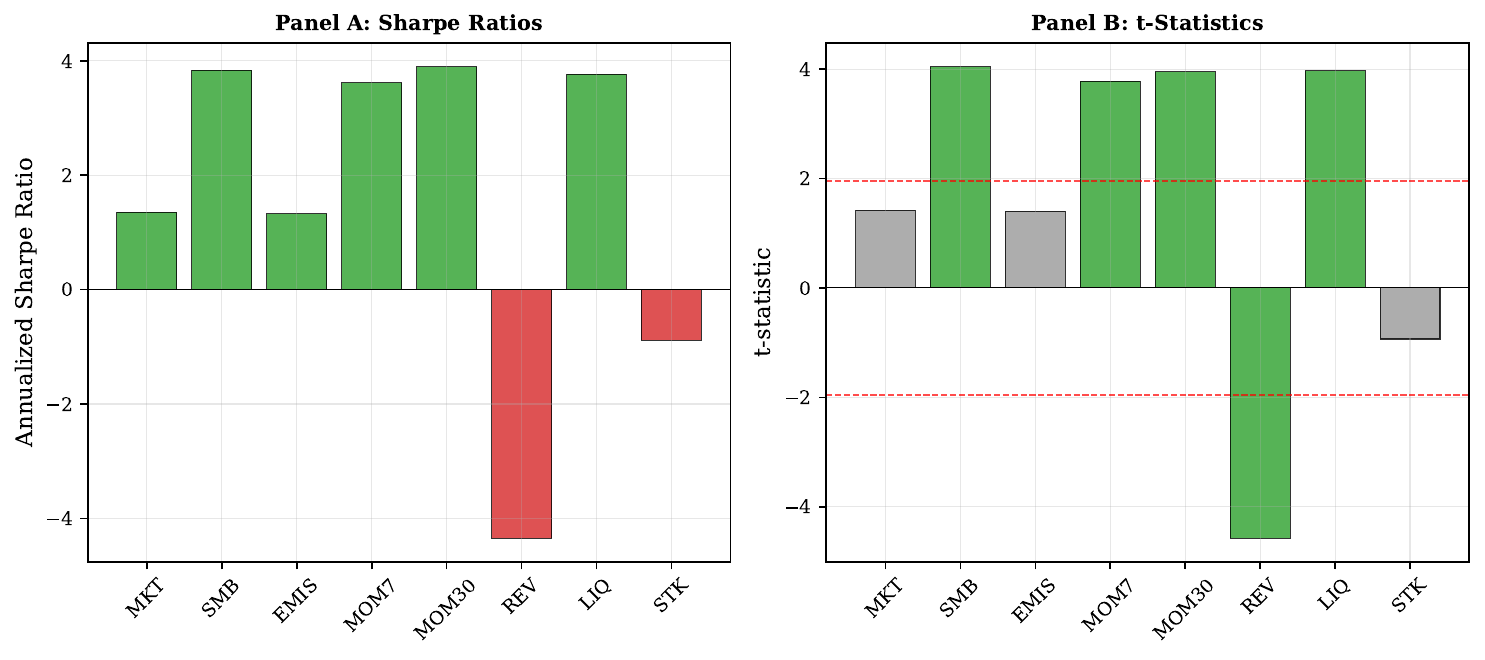}
\caption{Factor Sharpe Ratios and $t$-Statistics}
\label{fig:sharpe_tstats}
\begin{minipage}{\textwidth}
\small
Panel A: annualized Sharpe ratios. Panel B: $t$-statistics for the null hypothesis that mean factor returns equal zero. Dashed red lines indicate the 5\% significance threshold ($|t| = 1.96$). Green bars exceed the threshold; gray bars do not.
\end{minipage}
\end{figure}

\begin{figure}[htbp]
\centering
\includegraphics[width=\textwidth]{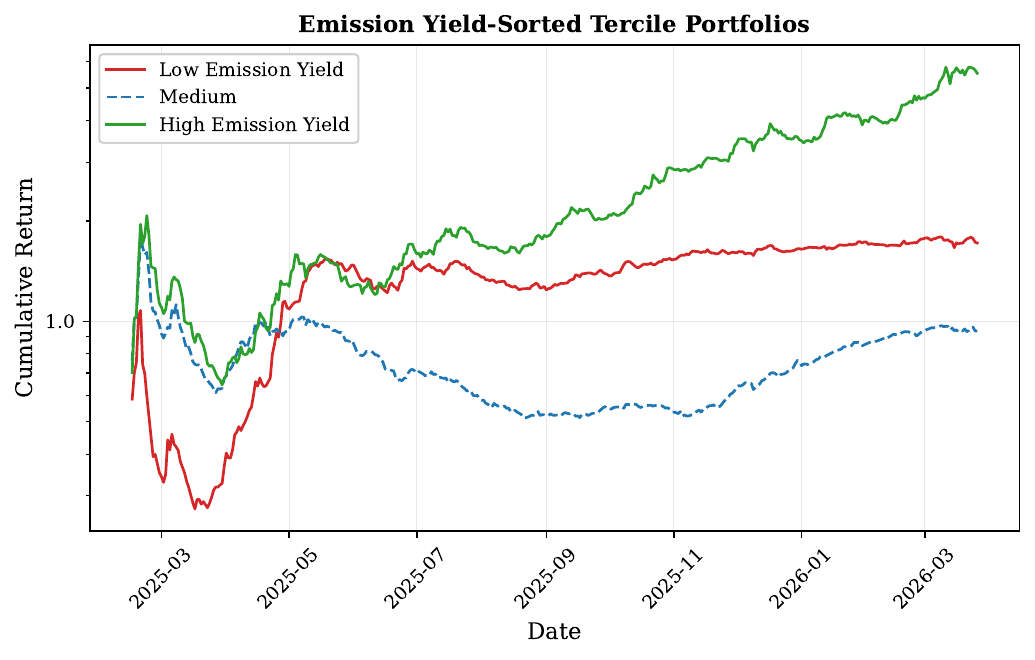}
\caption{Cumulative Returns of Emission Yield-Sorted Tercile Portfolios}
\label{fig:emission_portfolios}
\begin{minipage}{\textwidth}
\small
Subnets are sorted daily into terciles based on lagged emission yield (daily emission rate divided by market capitalization). High-emission-yield subnets outperform low-emission-yield subnets, analogous to the value premium in equities. Returns are in TAO terms, log scale.
\end{minipage}
\end{figure}

\begin{figure}[htbp]
\centering
\includegraphics[width=\textwidth]{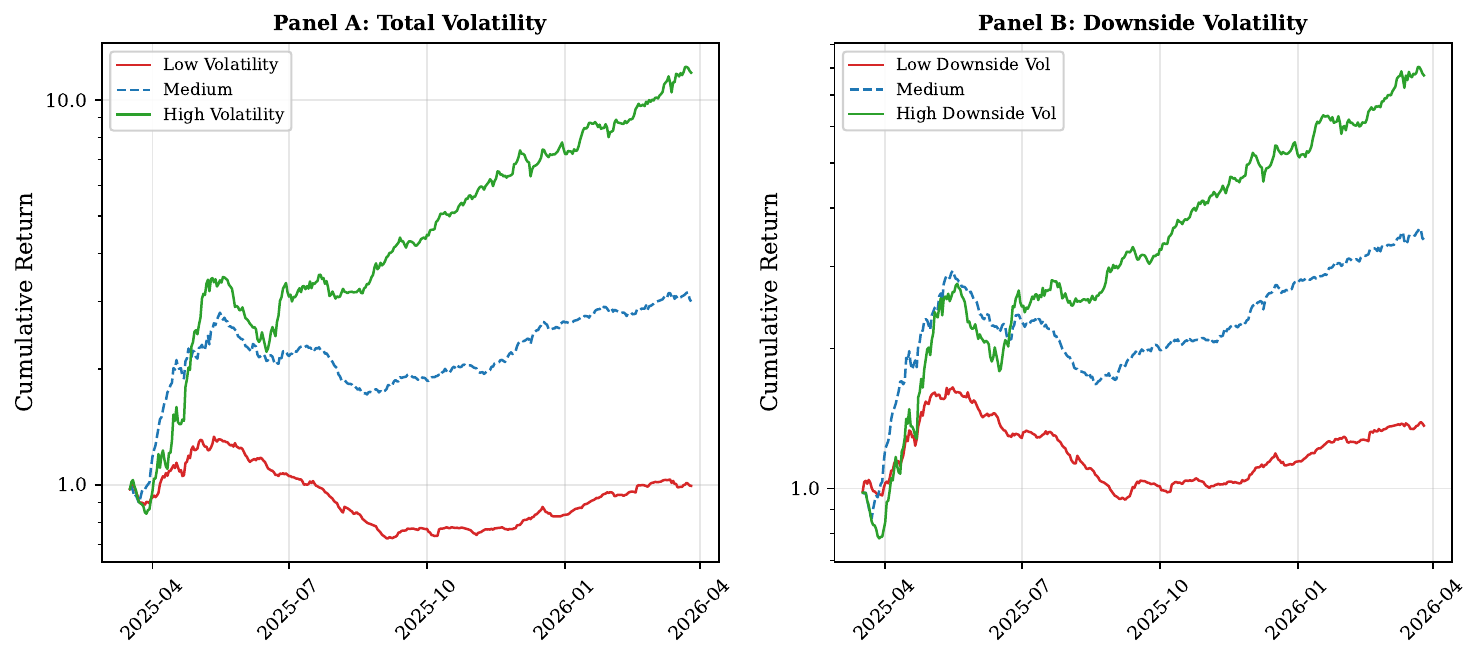}
\caption{Cumulative Returns of Volatility-Sorted Tercile Portfolios}
\label{fig:vol_portfolios}
\begin{minipage}{\textwidth}
\small
Panel A sorts subnets daily into terciles based on 30-day realized volatility. Panel B sorts on 30-day downside semi-deviation. In both cases, high-risk subnets outperform low-risk subnets, the opposite of the low-volatility anomaly in equities. The pattern reflects the mechanical link between size and volatility in constant-product AMMs.
\end{minipage}
\end{figure}

\begin{figure}[htbp]
\centering
\includegraphics[width=\textwidth]{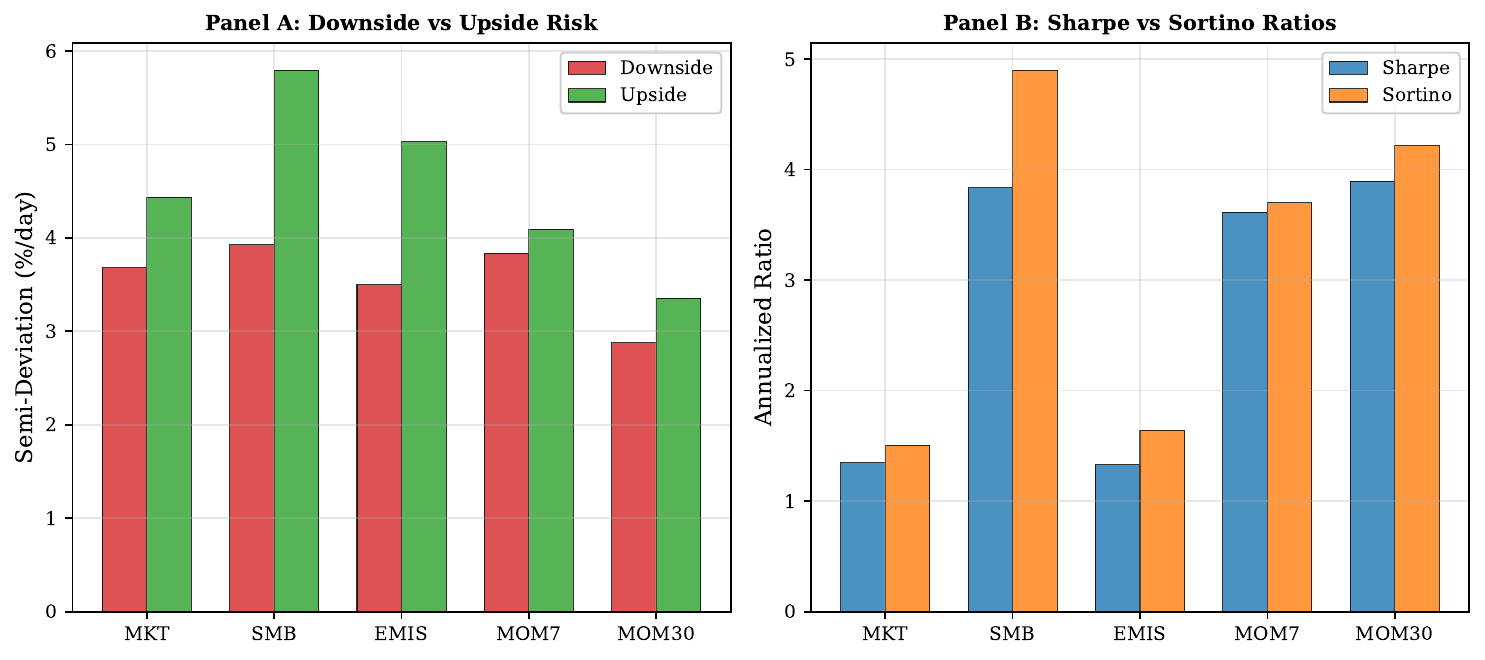}
\caption{Factor Return Risk Decomposition: Downside vs Upside and Sharpe vs Sortino}
\label{fig:risk_decomp}
\begin{minipage}{\textwidth}
\small
Panel A: downside and upside semi-deviations for each factor. SMB has favorable asymmetry (downside/upside ratio = 0.68). Panel B: annualized Sharpe and Sortino ratios. SMB's Sortino (4.90) exceeds its Sharpe (3.84), indicating that its risk-adjusted performance improves when measured against downside risk alone.
\end{minipage}
\end{figure}

\clearpage

\begin{thebibliography}{99}

\bibitem[Adams, Zinsmeister, and Robinson(2020)]{AdamsZinsmeisterSalem2020}
Adams, Hayden, Noah Zinsmeister, and Dan Robinson, 2020, Uniswap v2 core, Technical report, Uniswap.

\bibitem[Amihud(2002)]{Amihud2002}
Amihud, Yakov, 2002, Illiquidity and stock returns: Cross-section and time-series effects, \textit{Journal of Financial Markets} 5, 31--56.

\bibitem[Ang, Chen, and Xing(2006)]{AngChenXing2006}
Ang, Andrew, Joseph Chen, and Yuhang Xing, 2006, Downside risk, \textit{Review of Financial Studies} 19, 1191--1239.

\bibitem[Asness, Moskowitz, and Pedersen(2013)]{AsnessMoskowitzPedersen2013}
Asness, Clifford S., Tobias J.\ Moskowitz, and Lasse Heje Pedersen, 2013, Value and momentum everywhere, \textit{Journal of Finance} 68, 929--985.

\bibitem[Bianchi and Babiak(2021)]{BianchiBabiak2022}
Bianchi, Daniele, and Mykola Babiak, 2021, A factor model for cryptocurrency returns, CERGE-EI Working Paper 710.

\bibitem[Capponi and Jia(2021)]{CapponiJia2021}
Capponi, Agostino, and Ruizhe Jia, 2021, The adoption of blockchain-based decentralized exchanges, Working paper, Columbia University.

\bibitem[Fama and French(1993)]{FamaFrench1993}
Fama, Eugene F., and Kenneth R.\ French, 1993, Common risk factors in the returns on stocks and bonds, \textit{Journal of Financial Economics} 33, 3--56.

\bibitem[Gibbons, Ross, and Shanken(1989)]{GRS1989}
Gibbons, Michael R., Stephen A.\ Ross, and Jay Shanken, 1989, A test of the efficiency of a given portfolio, \textit{Econometrica} 57, 1121--1152.

\bibitem[Fama and MacBeth(1973)]{FamaMacBeth1973}
Fama, Eugene F., and James D.\ MacBeth, 1973, Risk, return, and equilibrium: Empirical tests, \textit{Journal of Political Economy} 81, 607--636.

\bibitem[Frazzini and Pedersen(2014)]{FrazziniPedersen2014}
Frazzini, Andrea, and Lasse Heje Pedersen, 2014, Betting against beta, \textit{Journal of Financial Economics} 111, 1--25.

\bibitem[Fieberg et~al.(2025)]{FiebergEtAl2025}
Fieberg, Christian, Gerrit Liedtke, Thorsten Poddig, Thomas Walker, and Adam Zaremba, 2025, A trend factor for the cross section of cryptocurrency returns, \textit{Journal of Financial and Quantitative Analysis} 60, 3116--3153.

\bibitem[Kyle(1985)]{Kyle1985}
Kyle, Albert S., 1985, Continuous auctions and insider trading, \textit{Econometrica} 53, 1315--1335.

\bibitem[Jegadeesh and Titman(1993)]{JegadeeshTitman1993}
Jegadeesh, Narasimhan, and Sheridan Titman, 1993, Returns to buying winners and selling losers: Implications for stock market efficiency, \textit{Journal of Finance} 48, 65--91.

\bibitem[Liu and Tsyvinski(2021)]{LiuTsyvinski2021}
Liu, Yukun, and Aleh Tsyvinski, 2021, Risks and returns of cryptocurrency, \textit{Review of Financial Studies} 34, 2689--2727.

\bibitem[Liu, Tsyvinski, and Wu(2022)]{LiuTsyvinskiWu2022}
Liu, Yukun, Aleh Tsyvinski, and Xi Wu, 2022, Common risk factors in cryptocurrency, \textit{Journal of Finance} 77, 1133--1177.

\bibitem[Lui and Sun(2025)]{LuiSun2025}
Lui, Elizabeth, and Jiahao Sun, 2025, Bittensor protocol: The Bitcoin in decentralized artificial intelligence? A critical and empirical analysis, arXiv:2507.02951.

\bibitem[Moskowitz, Ooi, and Pedersen(2012)]{MoskowitzTeoGrinblatt2012}
Moskowitz, Tobias J., Yao Hua Ooi, and Lasse Heje Pedersen, 2012, Time series momentum, \textit{Journal of Financial Economics} 104, 228--250.

\bibitem[Rao(2020)]{Rao2020}
Rao, Yuma, 2020, Bittensor: A peer-to-peer intelligence market, Technical report, Opentensor Foundation.

\bibitem[Steeves et~al.(2022)]{Steeves2022}
Steeves, Jacob, Ala Shaabana, Yuqian Hu, Francois Luus, Sin Tai Liu, and Jacqueline Dawn Tasker-Steeves, 2022, Incentivizing intelligence: The Bittensor approach, Workshop paper, Decentralization and Trustworthy Machine Learning in Web3, NeurIPS 2022.

\end{thebibliography}
\end{document}